\documentclass[12pt,a4paper,english,superscriptaddress,aps,nofootinbib]{revtex4}
\usepackage[utf8]{inputenc}
\usepackage[T1]{fontenc}
\usepackage{amsmath,amssymb,graphicx}
\makeatletter
\usepackage{babel}
\usepackage[active]{srcltx}
\usepackage{graphicx,color}
\usepackage{changebar}
\usepackage{hyperref}
\hypersetup{
	colorlinks=true,
	urlcolor= blue,
	citecolor=blue,
	linkcolor= blue,
	bookmarks=true,
	bookmarksopen=false,}

\usepackage[T1]{fontenc}
\usepackage{ae}
\usepackage{amsmath}
\usepackage{braket}
\usepackage{mathtools}
\usepackage{slashed}
\usepackage{empheq}
\usepackage{tikz}
\usepackage{multirow}
\usetikzlibrary{decorations.pathmorphing}
\usepackage[caption = false]{subfig}

\usepackage{graphicx}
\usepackage{amsfonts}
\usepackage{amsmath}
\newcommand{\be}{\begin{equation}}
	\newcommand{\ee}{\end{equation}}
\newcommand{\bea}{\begin{eqnarray}}
	\newcommand{\eea}{\end{eqnarray}}

\begin{document}	
 
 \title{Effects of quantum fluctuations of the metric on a braneworld}



\author{C. A. S. Almeida}
\email{carlos@fisica.ufc.br}
\affiliation{Institute of Cosmology, Department of Physics and Astronomy, Tufts University, Medford, Massachusetts, USA.}
\affiliation{Departamento de F\'{\i}sica, Universidade Federal do Cear\'{a}, Fortaleza, CE, Brazil.}

\author{F. C. E. Lima}
\email{cleiton.estevao@fisica.ufc.br}
\affiliation{Departamento de F\'{i}sica, Universidade Federal do Maranh\~{a}o, S\~{a}o Lu\'{i}s, MA,  Brazil.}

\begin{abstract}
\vspace{0.5cm}

\noindent \textbf{Abstract:} Adopting the premise that the expected value of the quantum fluctuating metric is linear, i.e., $\langle g^{\mu\nu}\rangle=\alpha g^{\mu\nu}$, we analyze the modified gravity theory induced by the Einstein-Hilbert action coupled to a matter field. This approach engenders the $f(R,T)$ gravity used to investigate the braneworld. In this scenario, considering a thick brane, the influence of metric fluctuations on brane dynamics is investigated. Consequently, one shows how the metric fluctuations influence the vacuum states. This influence has repercussions for modifying the brane energy and the asymptotic profile of the matter field. After noticing these modifications, we analyzed the most likely and stable structures from the matter field. One performs this analysis considering the theoretical measure of differential configurational entropy.

\end{abstract}

\maketitle

\thispagestyle{empty}

\newpage 

\section{Introduction}

A challenge to theoretical gravity models is their agreement with recent observational data. For example, some data suggest the existence of a late acceleration from the universe \cite{Harko} and the possibility of existing matter and dark energy on it \cite{Riess,Perlmutter,Bernadis,Hanany}. In this case, the proposals that allow good agreement between theoretical models and phenomenological data are the modified gravity theories \cite{Harko,Sotiriou1,Sotiriou2,Nojiri,Felice,Sharif}. Physically, one assumes in the modified gravity models that Einstein's gravity of general relativity is modified to formulate a more general action \cite{NojiriSDO}. The simplest possibilities for constructing a modified gravity theory are the models $f(R)$ \cite{Felice,Starobinsky,Olmo1,Olmo2,Olmo3}, $f(T)$ \cite{YFCai,SHChen,Dent,RJYang}, and $f(R, T)$ \cite{Harko,Barrientos,JBarrientos,Alvarenga,Moraes}, which briefly are theories whose Einstein-Hilbert standard action is replaced by an arbitrary function of the Ricci scalar $f(R)$ or by the trace of the stress-energy tensor $f(T)$ or simultaneously by the Ricci scalar $R$ and the trace of the stress-energy tensor $T$, i. e., a $f(R, T)$ gravity. Thus, motivated by this, several studies have considered the modified gravity theory, see Refs. \cite{ADas,Myrzakulov,RZaregonbadi,ZYousaf}.

Naturally, a question arises when studying the modified gravity theories. That issue is: what proposal is the most adequate to describe the modified gravity theory? Some works propose specific models of modified gravity considering weak field constraints obtained in the classical tests of general relativity for models similar to the solar system \cite{TChiba1,ALErickcek,TChiba2,NojiriSDO00,CapozzielloST1,CapozzielloST2}. However, in this work, we will consider a different approach, i.e., we will apply a quantum fluctuations approach of the metric in the Einstein-Hilbert theory to obtain a modified gravity theory. The quantum fluctuation approach to the metric entails considering a system within quantum gravity where the metric comprises a classical contribution ($g^{\mu\nu}$) and a quantum contribution ($\delta\hat{g}^{\mu\nu}$). In contemplating this approach, it is interesting to note that the contribution from the expected value of the quantum part must be non-null. This consideration arises from the role played by the quantum component in the metric's decomposition, which engenders modifications to Einstein's gravity structure, thereby enabling the emergence of a non-minimal coupling between gravity and matter. For this approach, the expected value of the metric adheres to certain physical assumptions, including the non-nullity of the expected value of the quantum contribution and the feasibility of expressing the metric $g^{\mu\nu}$ in terms of some symmetric second-order tensor constructed from the metric. Naturally, a more fundamental scenario that respects these impositions is when the expected value of the metric exhibits a linear behavior, i.e., $\langle \hat{g}^{\mu\nu}\rangle=\alpha g^{\mu\nu}$. As a consequence of this approach, one notes that the small fluctuations induce linear $f(R, T)$ theories. This result will allow us to study the impacts of the quantum fluctuations of metric on the braneworld scenario in five dimensions.

Theoretical models that consider the existence of extra dimensions start from the premise that the universe is in a higher-dimensional spacetime \cite{Rubakov,Koyama,Rubakov2}. Considering this, these theories have attracted the attention of several researchers. An interesting theory in this scenario is the braneworld theory \cite{Ida,Tan,Gauy,Germani,Sahni}. The idea of braneworld begins to gain supporters with the proposal of the Horava-Witten theory, which relates the heterotic string theory $\mathrm{E}_8\times\mathrm{E}_8$ coupled to $M$ theory with eleven-dimensions compacts \cite{Koyama}. In this scenario, supergravity lives at the 5-dimensional Anti-de-Sitter (AdS) spacetime. Meantime, Standard Model particles are confined to the 3-brane \cite{Koyama}. Phenomenologically, this hypothesis opens a way to resolve the mass hierarchy problem between the fundamental scales of particle physics and gravity. Based on this, Randall and Sundrum \cite{RS1,RS2} proposed the five-dimensional braneworld model that became known by their name. In his theory, one assumes the existence of four-dimensional domain walls contained in a five-dimensional AdS spacetime \cite{RS1,RS2}.

The braneworld models on modified gravity scenarios have been a topic of increasing interest, see Refs. \cite{Wang,Saavedra,Hoff,Guo2}. Generally speaking, one believes that these braneworld models in modified gravity scenarios can provide sophisticated solutions to hierarchy problems, further answering some issues about the description of dark matter \cite{Aspeitia} and dark energy \cite{Koyama2}, as well as other questions \cite{Brax,Alam}. Due to this, braneworld theories have gained space in some investigations \cite{Biggs,Pourhassan,Sokoliuk}. That is because, in this scenario, one can more easily notice the effects of modified gravity on the brane, see Ref. \cite{MSLA}. Thus, motivated by this, we will investigate a thick brane scenario in a $f(R, T)$ gravity, seeking to understand how the metric fluctuations influence a five-dimensional braneworld.

Not far from these theories, we will consider the Configurational Entropy (CE) initially proposed by Gleiser et al. \cite{Gleiser1,Gleiser2,Gleiser3,Gleiser4,Gleiser5,Gleiser6} to find the values of the metric fluctuations that describe the most likely and stable braneworld. Thus, we hope to obtain the most likely behavior of the brane in an $f(R,T)$ gravity theory induced by quantum fluctuations from the metric. Indeed, to reach these results, let us adopt a variant of CE, i.e., the Differential Configurational Entropy (DCE). We will do this because DCE has proven appropriate in studying the localized structures that arise in braneworld theories, e. g., see Ref. \cite{Roldao4}. Moreover, the DCE is a good approach in investigations from models that admit topological defects type domain walls \cite{Lima3}. That is because DCE can provide adequate information about the parameters that describe a stable field configuration \cite{Gleiser1,Gleiser2,Gleiser3,Gleiser4,Gleiser5,Gleiser6}. Examples of the application of this approach appear in Ref. \cite{Gleiser2}, where the authors show that the energy variations of the structures are proportional to theoretical measures of CE and its variants. Furthermore, this approach has reported significant results on the dynamics of spontaneous symmetry breaking \cite{Gleiser1}, of compact objects \cite{Gleiser3,Gleiser7}, and the stability of modified gravity models on braneworlds \cite{FMoreira1,FMoreira2}.

Based on all the applications and concepts presented throughout this introduction, the question naturally arises: what is the influence of metric fluctuations in a modified gravity scenario? Also, how do these fluctuations are felt at the braneworld? In this paper, let us answer these questions.

We organized our work as follows: in section II, one considers the quantum fluctuations approach from the metric to induce an f(R) gravity. In section III, one builds a braneworld theory using the modified gravity theory. In section IV, we adopt the approach of configurational entropy to select the most likely and stable regimes associated with the braneworld. To finalize, in section V our discoveries are announced.

\section{Quantum fluctuations inducing a modified gravity}

A major open problem in theoretical physics is the quantization of gravity. In this scenario, to quantize gravity, several steps are required. The first step towards a quantized theory of gravity is to assume that metric is an ordinary field. Thus, in this section, let us adopt this premise for the metric profile. Indeed, this will allow, in principle, us to build a gravity-effective theory. To apply this approach, we employ a metric fluctuation on the Einstein-Hilbert Lagrangian density, i. e.,
\begin{align}\label{Lagrangian}
    \mathcal{L}= \sqrt{-g}\bigg[-\frac{1}{4}R+\mathcal{L}_{\text{matter}}\bigg]
\end{align}
where the Lagrangian density of the matter field is
\begin{align}\label{LMatter}
    \mathcal{L}_{\text{matter}}=\frac{1}{2}\nabla_\mu\phi\,\nabla^\mu\phi-V(\phi).
\end{align}

To apply a non-perturbative quantization approach, let us promote the classical fields to field operators. Thus, Einstein's equation takes the form: 
\begin{align}\label{EqEinstein}
    \hat{G}_{\mu\nu}=\hat{R}_{\mu\nu}-\frac{1}{2}g_{\mu\nu}\hat{R}=\kappa^2\hat{T}_{\mu\nu}.
\end{align}
Indeed, one can find a similar approach in Refs. \cite{Dzhunushaliev,Dzhunushaliev1}. For example, in Ref. \cite{Dzhunushaliev}, one shows that the metric in a gravity quantum extension has classical and quantum contributions. Meanwhile, Dzhunushaliev \cite{Dzhunushaliev1} et al. apply this approach to explain the acceleration of the Universe.

A feature of this approach is that the quantities $\Gamma^{\rho}\,_{\mu\nu}$, $R^{\rho}\,_{\lambda\mu\nu}$ and $R_{ \mu\nu}$ are initially promoted to operators so that their classic definitions are not changed. Thus, assuming these quantities as operators $\hat{\Gamma}^{\rho}\,_{\mu\nu}$, $\hat{R}^{\rho}\,_{\lambda\mu\nu}$ and $\hat{R}_{\mu\nu}$ one can considering Heisenberg's formalism, solve the equation of operator (\ref{EqEinstein}) by averaging over all possible products of the metric operator $\hat{g}(x_i)$. That gives us an infinite set of equations, namely,
\begin{align}\label{QuantEq}
    \begin{matrix}
        \langle\mathcal{Q}\vert\hat{g}(x_1)\cdot\hat{G}_{\mu\nu}\vert\mathcal{Q}\rangle=\kappa^2\langle\mathcal{Q}\vert\hat{g}(x_1)\cdot\hat{T}_{\mu\nu}\vert\mathcal{Q}\rangle\\
        \langle\mathcal{Q}\vert\hat{g}(x_1)\cdot\hat{g}(x_2)\cdot\hat{G}_{\mu\nu}\vert\mathcal{Q}\rangle=\kappa^2\langle\mathcal{Q}\vert\hat{g}(x_1)\cdot\hat{g}(x_2)\cdot\hat{T}_{\mu\nu}\vert\mathcal{Q}\rangle\\
        \vdots\\
        \langle\mathcal{Q}\vert\hat{g}(x_1)\cdot ...\cdot\hat{g}(x_i)\hat{G}_{\mu\nu}\vert\mathcal{Q}\rangle=\kappa^2\langle\mathcal{Q}\vert\hat{g}(x_1)\cdot ...\cdot\hat{g}(x_i)\cdot\hat{T}_{\mu\nu}\vert\mathcal{Q}\rangle .
    \end{matrix}
\end{align}
In this scenario, we interpret $\vert\mathcal{Q}\rangle$ as a quantum state. An approach to solving the system of Eq. (\ref{QuantEq}) is to decompose the metric operator as follows:
\begin{align}\label{fluctuation}
    \hat{g}_{\mu\nu}=g_{\mu\nu}+\delta \hat{g}_{\mu\nu},
\end{align}
so that $g_{\mu\nu}$ is the average metric and $\delta\hat{g}_{\mu\nu}$ is a fluctuation of the non-trivial metric. For more details, see Refs. \cite{Dzhunushaliev,Dzhunushaliev1,Dzhunushaliev2}. Using the perturbation (\ref{fluctuation}), let us expand the Lagrangian (\ref{Lagrangian}). In this case, disregarding higher-order perturbations, one obtains
\begin{align}
    \mathcal{L}\approx-\frac{1}{4}\sqrt{-g}(R+G_{\mu\nu}\langle\hat{g}^{\mu\nu}\rangle)+\sqrt{-g}\bigg(\mathcal{L}_{\text{matter}}+\frac{1}{2}T_{\mu\nu}\langle\delta g^{\mu\nu}\rangle\bigg).
\end{align}

As suggested by Yang in Ref. \cite{RYang}, it is convenient to assume linear metric fluctuations due to symmetries of the Lagrangian density (\ref{Lagrangian}). Thereby, adopting a linear metric fluctuation, i.e., $\langle\hat{g}^{\mu\nu}\rangle\approx \alpha g^{\mu\nu}$ \footnote{We adopt the expected value of the linear metric $\langle g^{\mu\nu}\rangle \approx \alpha g^{\mu\nu}$ once our interest consists in developing a $f(R, T)$ gravity. However, the constraint on $\langle g^{\mu\nu}\rangle$ are $\langle g^{\mu\nu}\rangle$ is non-zero and a symmetric two-rank tensor. Thus, one has several possibilities a priori. For instance, adopting $\langle g^{\mu\nu}\rangle \simeq \alpha R^{\mu\nu}$, one obtains a quadratic gravity theory, see Refs. \cite{Dzhunushaliev,Dzhunushaliev1}.}, we obtain the modified Lagrangian density, namely, 
\begin{align}\label{Lagrangian2}
    \mathcal{L}=\sqrt{-g}\bigg[-\frac{1}{4}(1-\alpha)R-\frac{1}{2}\alpha T+\mathcal{L}_{\text{matter}}\bigg].
\end{align}
In this formulation, $\alpha$ denotes a dimensionless parameter responsible for adjusting the quantum fluctuations of the metric. Seeking a theory that approaches the usual theory, i.e., the Einstein-Hilbert theory, let us assume that the parameter $\alpha$ describes the small fluctuations of the metric, consequently, $\vert\alpha\vert\ll 1$. Furthermore, the $R$ displayed on the action is the Ricci scalar, and $T$ is the trace of the stress-energy tensor $T=g^{\mu\nu}T_{\mu\nu}$, so that the stress-energy tensor is
\begin{align}
    T_{\mu\nu}=-\frac{2}{\sqrt{-g}}\frac{\delta\sqrt{-g}\mathcal{L}_{\text{matter}}}{\delta g^{\mu\nu}}.
\end{align}

\section{The braneworld in $f(R,T)$ gravity}

Assuming the corrections arising from the metric fluctuations, let us build a braneworld considering a modified gravity scenario type $f(R,T)=\lambda(R)+\xi(T)$, i. e.,
\begin{align}\label{MAction1}
    S=\int\, d^5x\, \bigg[-\frac{1}{4}\lambda(R)-\xi(T)+\mathcal{L}_{\text{matter}}\bigg]
\end{align}
with
\begin{align}\label{choise}
    \lambda(R)=(1-\alpha)R \hspace{1cm} \text{and} \hspace{1cm} \xi(T)=\frac{1}{2}\alpha T.
\end{align}
The $\lambda(R)$ and $\xi(T)$ functions chosen in Eq. (\ref{choise}) is the result presented in the Lagrangian (\ref{Lagrangian2}). Furthermore, we will adopt $4\pi G_5=1$ and $g=$det$(g_{ab})$.

To study the braneworld in $f(R,T)$ gravity, allow us to assume the line element
\begin{align}
    ds^2=\text{e}^{2A(y)}\eta_{ab}dx^a dx^b-dy,
\end{align}
where e$^{2A}$ is called the warp factor. Here, the indices $a$ and $b$ are varying from 0 to 3 with the extra-dimension being $y=x^4$ and metric signature $(+,-,-,-,-)$.

Being the matter field described by Lagrangian density (\ref{LMatter}), the stress-energy tensor will be 
\begin{align}\label{EnergyStressT}
    T_{\mu\nu}=\nabla_\mu\phi\,\nabla_\nu\phi-g_{\mu\nu}\mathcal{L}_{\text{matter}} \hspace{1cm} \text{with} \hspace{1cm} \mu,\nu=0,1,\dots,4;
\end{align}
so that the trace of the stress-energy tensor is
\begin{align}
    T=g^{\mu\nu} T_{\mu\nu}=-\frac{3}{2}\nabla_\mu\phi\,\nabla^\mu\phi+5V(\phi).
\end{align}

Let us now investigate the equations of motion of our five-dimensional braneworld in $f(R,T)$ gravity. For this, we start by varying the action concerning the scalar field $\phi$. That leads us to
\begin{align}\label{EqPhiF}
    \nabla_\mu\nabla^\mu\phi+3\nabla_\mu[\xi_T\nabla^\mu\phi]+(5\xi_T+1)V_\phi=0,
\end{align}
where $\xi_T=d\xi/dT$ and $V_\phi=dV/d\phi$.

Subsequently, varying the action (\ref{MAction1}) concerning the metric, one arrives at Einstein's equation, namely,
\begin{align}\label{EqEinst}
    \lambda_R R_{\mu\nu}-\frac{1}{2}g_{\mu\nu}\lambda+(g_{\mu\nu}\square-\nabla_\mu\nabla_\nu)\lambda_R=2T_{\mu\nu}+2g_{\mu\nu}\lambda+6\lambda_T \nabla_\mu\phi\,\nabla_\nu\phi
\end{align}
where $\lambda_R=d\lambda/dR$ with $R=20A'(y)+8A''(y)$. Here the prime notation refers to the derivative concerning the extra-dimension coordinate. 

Exposing the Eqs. (\ref{EqPhiF}) and (\ref{EqEinst}) in terms of the matter field $\phi$ and the warp function $A(y)$, one obtains:
\begin{align}\label{MEq1}
    \bigg(1+\frac{3}{2}\alpha\bigg)\phi''+4\bigg(1+\frac{3}{2}\alpha\bigg)A'\phi'=\bigg(1+\frac{5}{2}\alpha\bigg)\frac{\partial V}{\partial\phi};&\\ \label{MEq2}
    (1-\alpha)A''=-\frac{2}{3}\bigg(1+\frac{3}{2}\alpha\bigg)\phi'^2;&\\
    \label{MEq3}
    3(1-\alpha)A'^2=\frac{1}{2}\bigg(1+\frac{3}{2}\alpha\bigg)\phi'^2- \bigg(1+\frac{5}{2}\alpha\bigg)V.&
\end{align}
Algebraically, one can reduce this set of equations to the expressions:
\begin{align}
    \label{MEq4}
    \phi''+4A'\phi'=\bigg(\frac{1+5\alpha/2}{1+3\alpha/2}\bigg)\frac{\partial V}{\partial\phi},
\end{align}
and 
\begin{align}
    \label{MEq5}
    A''+4A'^2=-\frac{4}{3}\bigg(\frac{1+5\alpha/2}{1-\alpha}\bigg) V.
\end{align}

 Furthermore, it is interesting to highlight that the braneworld described by equations (\ref{MEq4}) and (\ref{MEq5}) has energy due to the propagation of the matter field along the extra-dimension. In this case, this energy is
\begin{align}\label{energy}
    E=\int\, \text{e}^{2A}\,\bigg[\frac{1}{2}\phi'^2+V(\phi)\bigg]\, dy,
\end{align}
where brane's energy density ($\rho_E$) is
\begin{align}\label{EnergyD}
    \rho_E(\phi; y)=\text{e}^{2A}\,\bigg[\frac{1}{2}\phi'^2+V(\phi)\bigg].
\end{align}

\subsection{The thick-brane model}

Allow us to continue our study assuming a particular geometry for spacetime. To choose a specific profile of spacetime, let us adopt an appropriate form for the warp function $A(y)$. Indeed, one bases this choice on some requirements. For example, in general, it is interesting to assume an $A(y)$ that reproduces a Randall-Sundrum type warp factor far from the brane, i.e., $\lim_{y\to\infty}\text{e}^{2A(y)}= 0$. Meanwhile, in the neighborhood of the brane, it should have a smooth profile (no singularity). That allows us to bypass the thin-brane energy scale problem and leads us to a thick-brane model. We found two distinct braneworld behaviors, i.e., the thin and thick brane. In both cases, one requests that the warp factor be symmetrical. Mathematically, $\text{e}^{2A(y)}$=$\text{e}^{2A(-y)}$ is required for the matter field to have $Z_2$ symmetry preserved, while the symmetry break occurs in the matter sector. Furthermore, another condition that restricts the behavior of the warp factor is the zero-mode normalization of the graviton $\int_{-\infty}^{\infty}\,dy\, \text{e}^{8A(y )}$ which must be finite and non-null. In addition to these requirements, let us assume a warp function profile that falls into the approximate thin-brane theory and bypasses the brane energy scaling problem. A warp function used that fulfills these requirements is the model in which the warp function takes the form:
\begin{align}\label{Afunction}
    A(y)=-\text{ln}[\cosh(\sigma y)].
\end{align}
In this context, the independent variable $y$ has a length dimension. Consequently, the parameter $\sigma$ is the inverse of length. So, the parameter $\sigma$ modifies the brane's width. Furthermore, one highlights that the warp function (\ref{Afunction}) has been used extensively in several models of braneworlds, see Refs. \cite{Moreira1,CCLi,TTSui,Moreira2}. For example, in Ref. \cite{Moreira1}, one assumes the function (\ref{Afunction}) to study braneworld theories in gravity $f(T, B)$. Meanwhile, adopting the warp function (\ref{Afunction}), the Ref. \cite{CCLi} presents an investigation of the effective theories with self-interacting. These applications are interesting because they show that the warp function of the type chosen in Eq. (\ref{Afunction}) accurately describes the behavior of the brane and can give us predictions about a thin-brane theory. That is possible because the brane thickness is adjustable by changing the $\sigma$ parameter.

\begin{figure}[ht!]
    \centering
    \includegraphics[height=7cm,width=8cm]{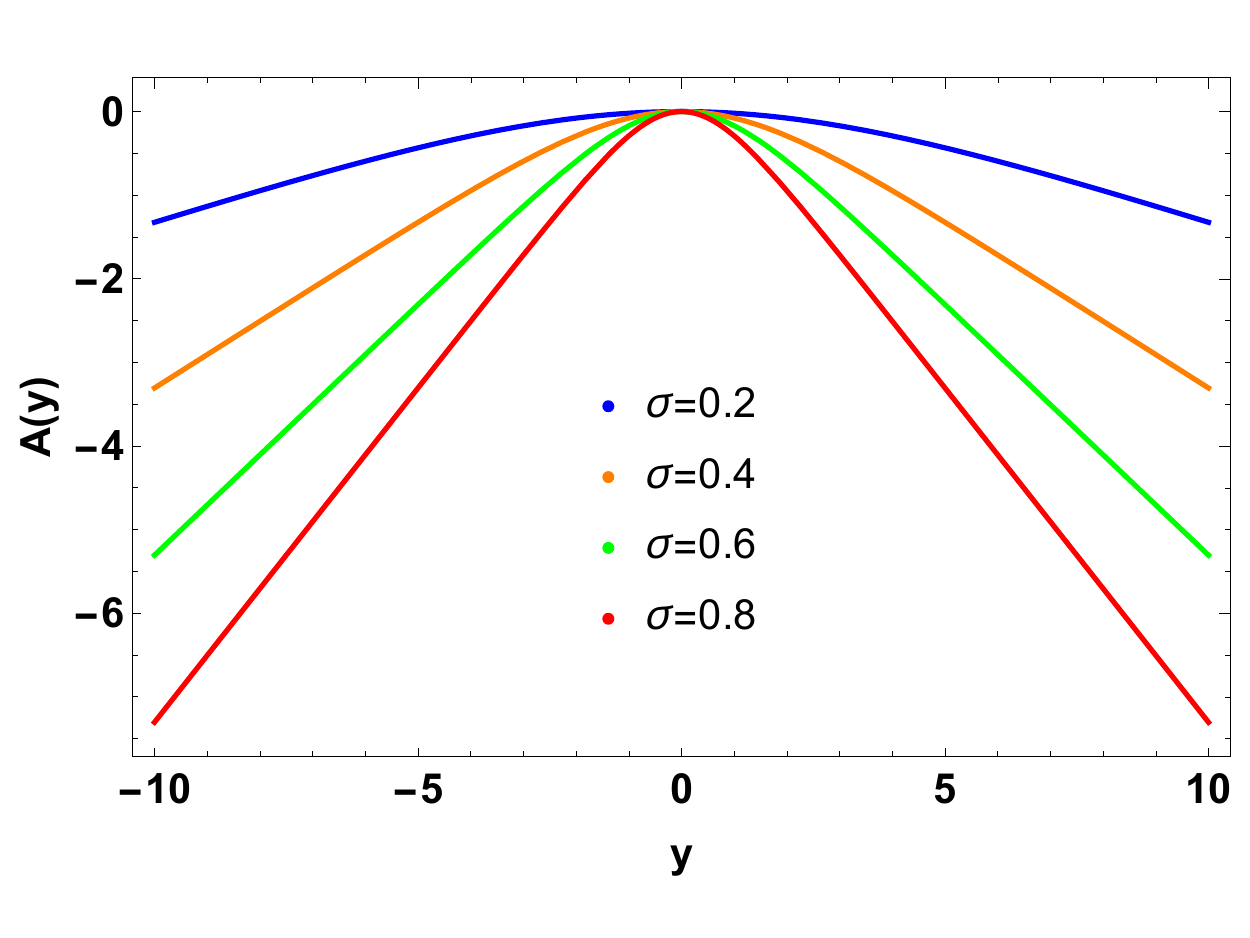}
    \includegraphics[height=7cm,width=8cm]{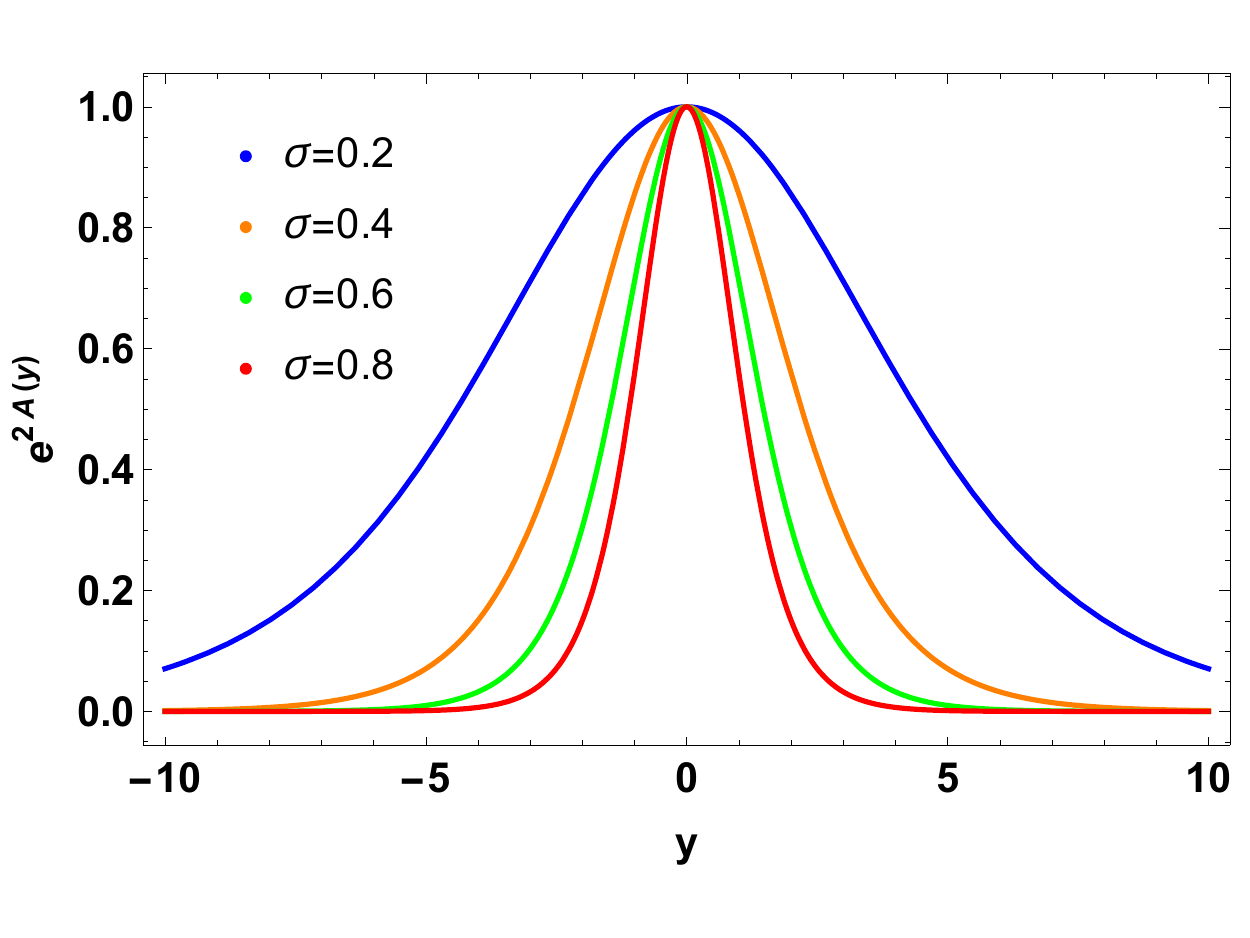}
    \vspace{-0.5cm}
    \begin{center}
      \hspace{0.3cm}  (a) \hspace{8cm} (b)
    \end{center}
    \vspace{-0.4cm}
    \caption{(a) Plots of the warp function $A(r)$. (b) The behavior of the warp factor e$^{2A(y)}$.}
    \label{fig1}
\end{figure}

We expose the behavior of the warp function $A(y)$ and the warp factor $\text{e}^{2A(y)}$ respectively in Figs. \ref{fig1}(a) and \ref{fig1}(b). Note that the warp function locates the brane at $y=0$ and reduces to zero over large distances, as required initially, but reproducing a thick brane behavior. However, for large values of $\sigma$, one obtains an effective theory of type thin brane. For convenience, in this study, we will consider the thick-brane case, i.e., when $\vert\sigma\vert\ll 1$.

Considering the profile of the function $A(y)$ (and consequently, the warp factor e$^{2A(y)}$) and Eq. (\ref{MEq2}), it is obtained that the solution of the matter field in terms of the extra dimension, the fluctuation parameters and the brane thickness. In this case, the matter field solution is
\begin{align}\label{MatterF}
    \phi(\sigma,\alpha;\,y)=2\sqrt{\frac{3(1-\alpha)}{2+3\alpha}}\arctan\bigg[\frac{\tanh(\sigma y)}{2}\bigg].
\end{align}
Meantime, considering the matter field solution and Eqs. (\ref{MEq4}) and (\ref{MEq5}), one concludes that the interaction is
\begin{align}\label{Pot}
    V(\sigma,\alpha;\,\phi)=&-\frac{3(1-\alpha)\sigma^2}{4+10\alpha}\bigg[4-5\, \text{sech}\bigg(2\, \text{arctanh} \bigg(\text{tan}\bigg(\frac{\phi}{2\sqrt{\frac{5}{2+3\alpha}-1}}\bigg)\bigg)\bigg)\bigg]
    \end{align}
with $\vert\phi\vert<\pi\sqrt{\frac{3(1-\alpha)}{2+3\alpha}}$.

\begin{figure}[ht!]
    \centering
    \includegraphics[height=7cm,width=8cm]{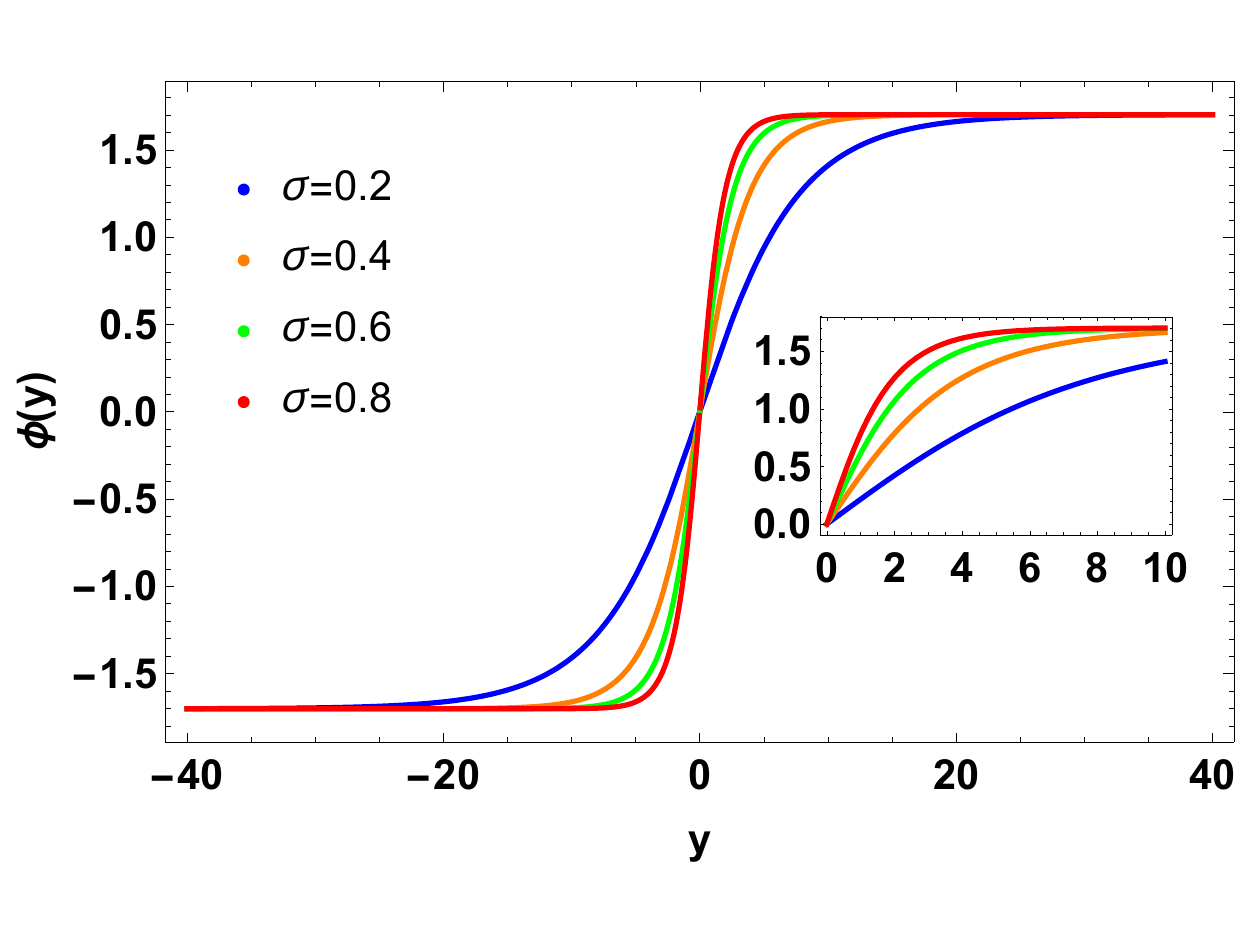}
    \includegraphics[height=7cm,width=8cm]{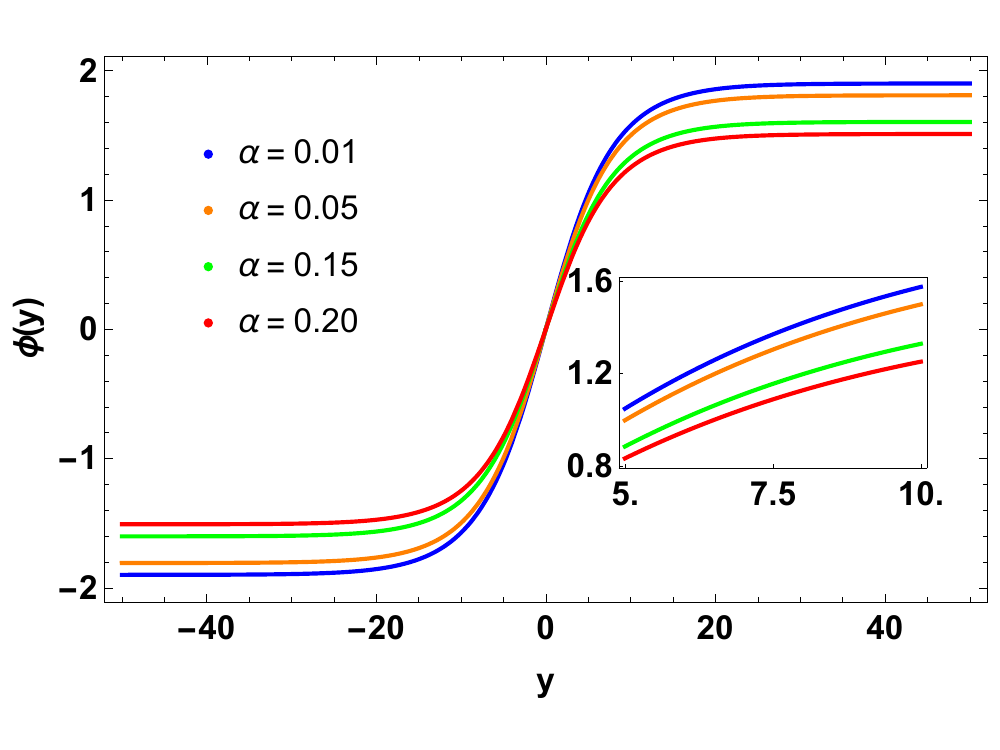}
    \vspace{-0.3cm}
    \begin{center}
      \hspace{0.3cm}  (a) \hspace{8cm} (b)
    \end{center}
        \vspace{-0.3cm}
    \caption{(a) Matter field as a function of the extra dimension for several values of the brane thickness and keeping $\alpha=0.1$. (b) Matter field as a function of the extra dimension for several values of the metric fluctuation and keeping $\sigma=0.2$.}
    \label{fig2}
\end{figure}

\begin{figure}[ht!]
    \centering
    \includegraphics[height=7cm,width=8cm]{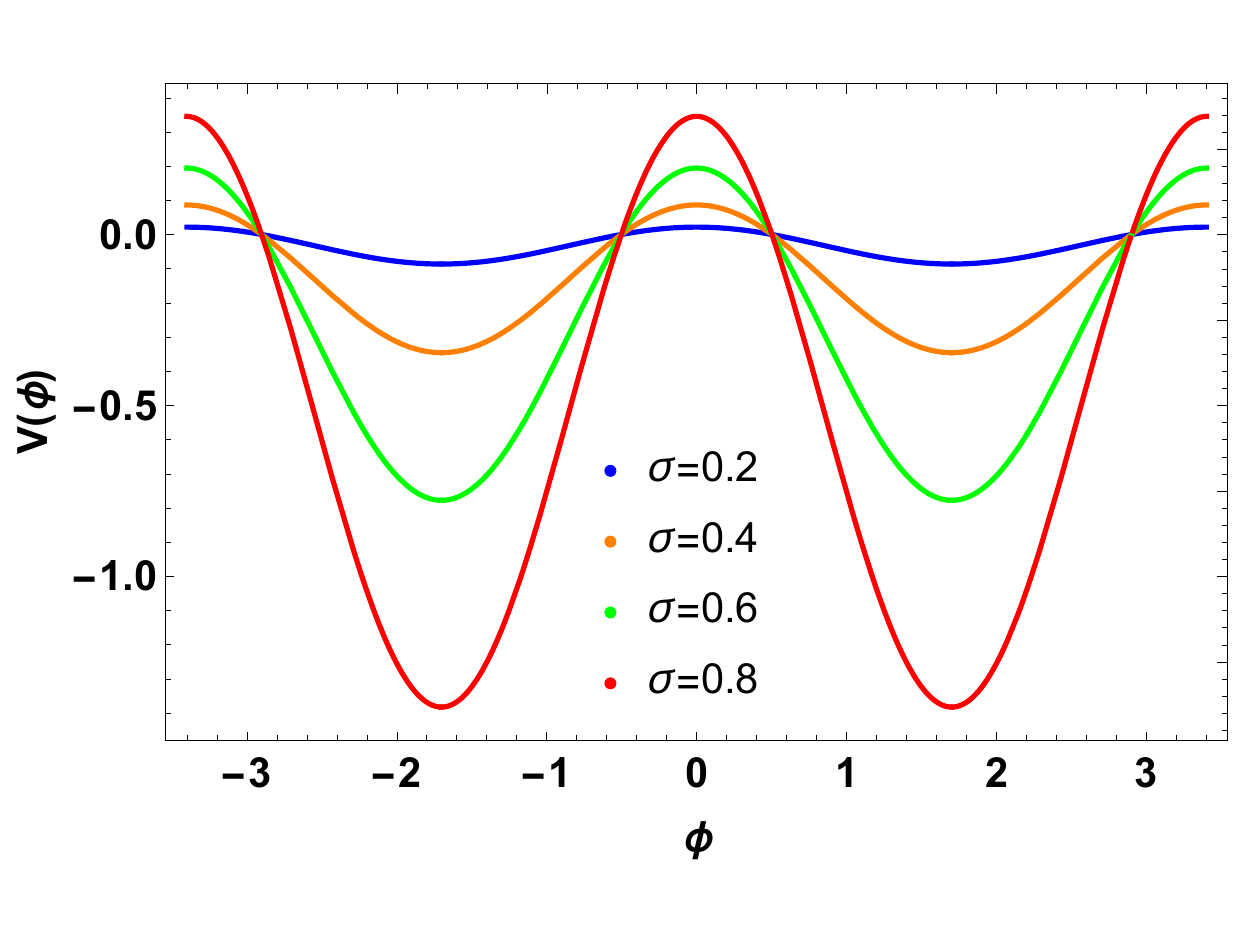}
    \includegraphics[height=7cm,width=8cm]{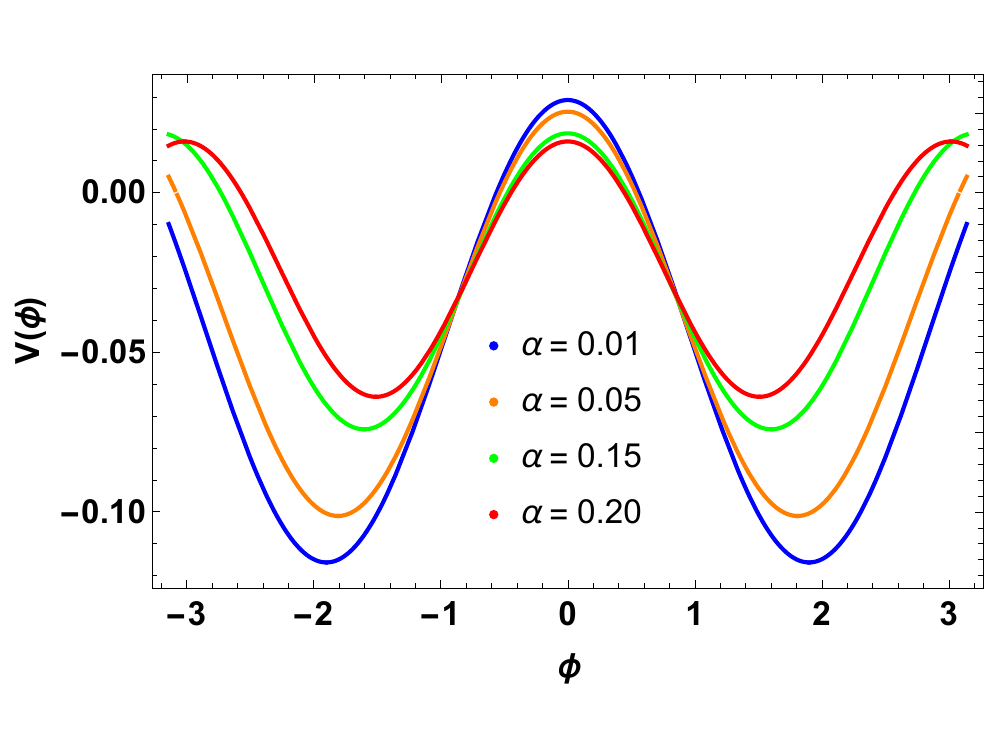}
    \vspace{-0.3cm}
    \begin{center}
      \hspace{0.3cm}  (a) \hspace{8cm} (b)
    \end{center}
    \vspace{-0.3cm}
    \caption{(a) Potential as a function of the matter field for several values of brane thickness and keeping $\alpha=10^{-1}$. (b) Potential as a function of the matter field for several values of the metric fluctuation and keeping $\sigma=0.2$.}
    \label{fig3}
\end{figure}

In Fig. \ref{fig2}(a), one displays the matter field for several values of the brane thickness. Moreover, Fig. \ref{fig2}(b) shows the behavior of the matter field for several values of the fluctuation parameter. Posteriorly, in Fig. \ref{fig3}, we expose the behavior of the interaction that satisfies Eqs. (\ref{MEq1}), (\ref{MEq2}) and (\ref{MEq3}) in terms of the field $\phi$.

The results found in Eqs. (\ref{MatterF} and \ref{Pot}) and shown in Figs. \ref{fig2} and \ref{fig3} show the field of matter described by solitonic solutions so that far from the brane, the theory reaches the vacuum of matter's topological sector. In fact, this vacuum value is $\phi_0=\pm\frac{\pi}{2}\sqrt{\frac{3(1-\alpha)}{2+3\alpha}}$. Therefore, the vacuum that arises due to spontaneous symmetry breaking in the matter sector modifies due to the metric fluctuations. Note that if the fluctuation of the metric reaches the value of $\alpha=1$, we will have $\phi_0=0$, and thus there will no longer be a vacuum. In other words, when $\alpha\to 1$, there will be no spontaneous symmetry breaking, so that matter field with symmetry $Z_2$ interpolating between the minimum energy configurations will not exist. Meantime, if the metric fluctuations are small, i.e., $\vert\alpha\vert<1$, spontaneous symmetry breaking is preserved. Thus, metric fluctuations will be perceived far from the brane regardless of the thickness $\sigma$. Besides, the results suggest that the variation in thickness will contract the matter field so that the thicker (greater the $\sigma$) the brane is, the smoother the topological transition of the matter field between the vacuums $ \phi_0$ of the theory. In contrast, if the brane tends to a thin brane-like behavior, i.e., $\vert\sigma\vert\ll 1$. Then, we will have contraction (or compactification) of the field of matter in a way that the matter field tends to evolve quickly into a vacuum.

\begin{figure}[ht!]
    \centering
    \includegraphics[height=7cm,width=8cm]{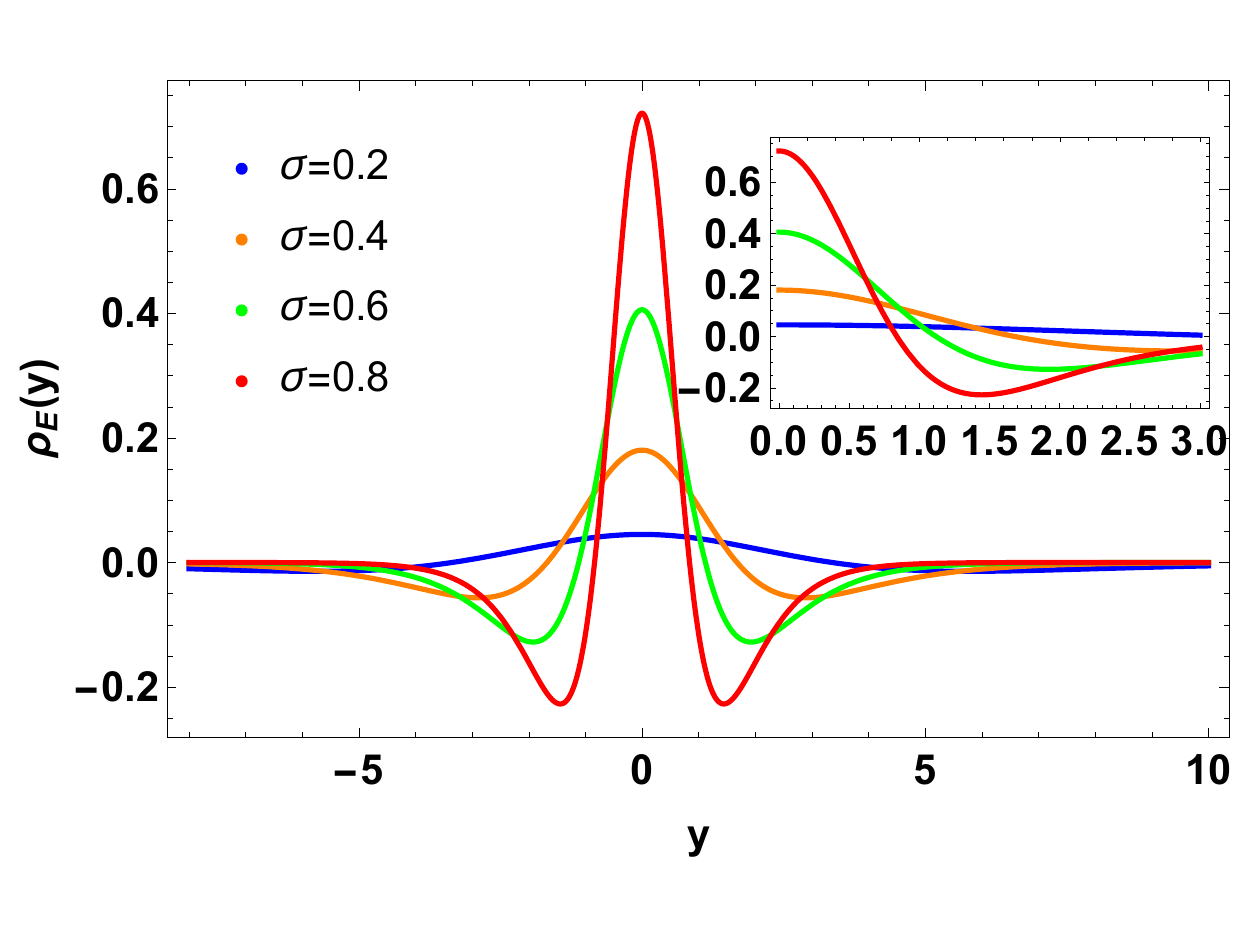}
    \includegraphics[height=7cm,width=8cm]{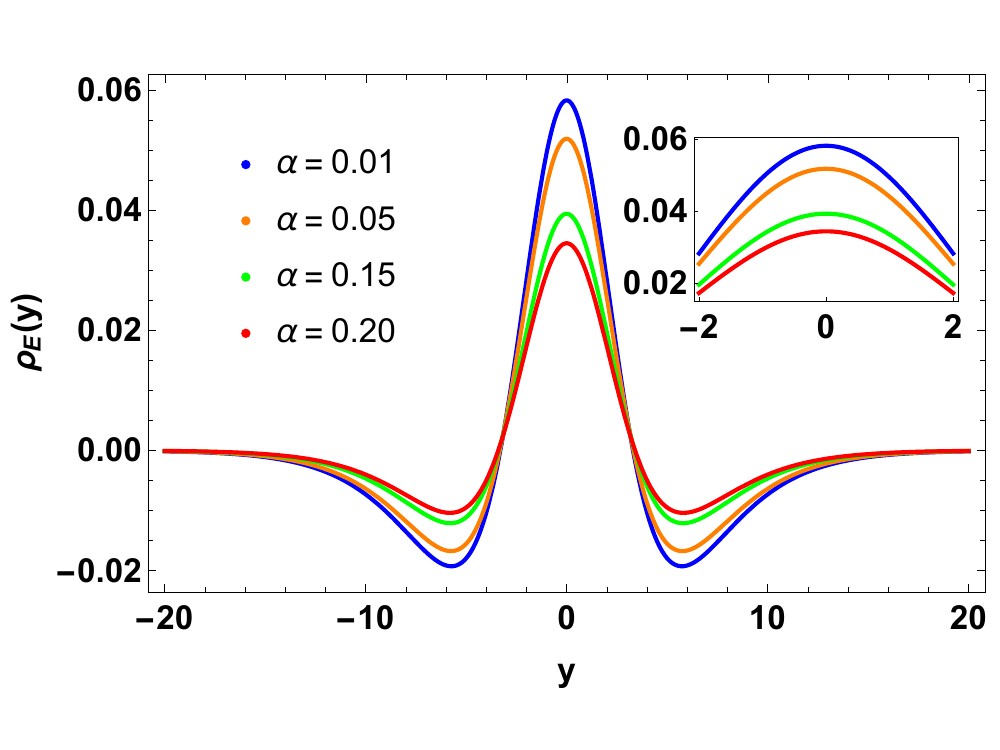}    
    \vspace{-0.3cm}
    \begin{center}
      \hspace{0.3cm}  (a) \hspace{8cm} (b)
    \end{center}
    \vspace{-0.3cm}
    \caption{(a) Energy density of the brane for several values of its thickness and keeping $\alpha=10^{-1}$. (b) Brane energy density for several values of the metric fluctuation and keeping $\sigma=0.2$.}
    \label{fig4}
\end{figure}

To finish this discussion of the topological brane theory, allow us to investigate the behavior of brane energy density. In this case, considering Eq. (\ref{EnergyD}) and the solutions found [Eqs. (\ref{MatterF}) and (\ref{Pot})], one obtains the brane energy density. We display the brane energy density in Figs. \ref{fig4}(a) and \ref{fig4}(b).

The brane energy density suggests a kink-like profile of the matter field. Also, as the quantum fluctuations of metric must be small, i.e., $\vert\alpha\vert\ll 1$, one notes the absence of internal structures in the thick brane when $0<\sigma<1$. As it was possible to predict, the brane has higher energy when the brane thickness decreases. This behavior occurs because there is a divergence when $\sigma\to 0$. So, to bypass this energy scaling problem, $\sigma$ must be contained in the range $0<\sigma\leq 1$. Moreover, we note that the modified gravity changes the vacuum value influencing the brane energy, i.e., when $\alpha$ decreases, the brane energy increases.

\section{Configurational information theoretical-measurement in braneworld in modified gravity}

The theoretical measure of information first appears with Claude E. Shannon in his seminal work on the mathematical theory of communication \cite{Shannon}. In his work, Shannon seeks to describe the best way to encode the information that an emitter transmits to a receiver. After Shannon's work, the definition of information entropy was reformulated and applied in several scenarios to obtain the information entropy in several theories, e.g., see to Refs. \cite{Gleiser1,Gleiser2,Gleiser3,Gleiser4,Gleiser5,Gleiser6,Roldao4,Roldao1,Roldao2,Roldao3,Roldao5}. Although Shannon's approach is suitable for investigating the information from systems of particles, in field theories, one needs to reformulate this due to the infinite degrees of freedom. Thus, based on Shannon's theory, Gleiser et al. \cite{Gleiser1,Gleiser2} propose an information entropy approach (or information theoretical measure) for the continuous limit \cite{Gleiser1}. This approach is also called Configurational Entropy (CE). In this scenario, the CE describes a representation of information-theoretic measures that detail the configurational complexity of the fields of a system \cite{Gleiser1,Gleiser2,Gleiser3,Gleiser4,Gleiser5}. In this work, we will use one variant of Configurational Entropy (CE), i.e., Differential Configurational Entropy (DCE). The use of this approach justifies by its applications. Indeed, the DCE has shown to be a good approach that gives us information about the informational content of the fields so that one can identify the most likely stable structures of the theory, see Refs. \cite{Roldao4,Correa1,Correa2,Correa3}. Motivated by the vast applications of DCE in the study of field theory \cite{Lima3,Lima1,Lima2}, in high energy physics \cite{Roldao6,Roldao7}, we are encouraged to adopt DCE to identify the most likely structures of our brane in $f(R,T)$ gravity.

Furthermore, it is pertinent to emphasize that the DCE formalism proves advantageous as it can provide appropriate values for the parameters governing domain walls \cite{LimaCasana}. In this section, by applying the principles of the DCE formalism, we will examine the set of values of the parameters $\alpha$ for generating the topological kinks that configure the matter sector of the braneworld in $f(R, T)$ gravity theory. It is worth noting that, in principle, the parameters of metric quantum fluctuations $\alpha$ can take any value, provided they are subject to the restriction $\vert\alpha\vert<1$. This analysis remains consistent once it allows for finding the most likely values for these parameters. Such inference is justified by the reduction in the complexity of the matter field as the DCE decreases, increasing the probability of obtaining specific field configurations \cite{Gleiser1,Gleiser2,Gleiser6}. Additionally, one employs this formalism in several studies of systems with localized energy configurations \cite{NBraga1,NBraga2,NBraga3}. Furthermore, it is pertinent to highlight that configurational entropy proves to be a good tool for studying the mass spectrum of kaon vector resonances \cite{RochaSilva}.

\subsection{Conceptual review of DCE applied to braneworld}

Before studying this theoretical measure of information from the braneworld in $f(R,T)$ gravity induced by metric fluctuations, let us start by presenting some concepts that underlie our study. To carry out the theoretical measurement of information, allow us to use the DCE concept. In this case, one defines the DCE in terms of Fourier's transform of the brane energy density, i.e.,
\begin{align}\label{FourierT}
    G[\omega]=\frac{1}{\sqrt{2\pi}}\int\, \rho_E(y)\,\text{e}^{i\omega y}\, dy.
\end{align}

Substituting Eq. (\ref{EnergyD}) into Fourier's transform (\ref{FourierT}), one obtains
\begin{align}\label{FourierT1}
    G[\omega]=\frac{1}{\sqrt{2\pi}}\int\, \bigg[\frac{1}{2}\phi'(y)^2+V(\phi(y))\bigg]\,\text{e}^{i\omega y+2A(y)}\, dy.
\end{align}

Considering Fourier's transform (\ref{FourierT1}), the modal fraction of the theory is constructed. This quantity is
\begin{align}\label{modalf}
    g(\omega)=\frac{\vert G(\omega)\vert^2}{\int\,\vert G(\omega)\vert^2\,d\omega}.    
\end{align}
By definition, the modal fraction is the weight relative to each wave mode at the reciprocal space. Thus, this quantity is always $\leq 1$. For more details, see Refs. \cite{Gleiser1,Gleiser2,Gleiser3,Gleiser4,Gleiser5,Gleiser6}.

Assuming the modal fraction defined in Eq. (\ref{modalf}), we define the DCE as
\begin{align}\label{DCE}
    S_C[g(\omega)]=-\int\,\bar{g}(\omega)\,\text{ln}[\bar{g}(\omega)]\,d\omega,
\end{align}
where the integrand of Eq. (\ref{DCE}) is called entropic density and $\bar{g}(\omega)$ is the modal fraction normalized.

Once defined the DCE, we are now ready to study the differential configurational entropy of the thick brane presented in the previous section.

\subsection{Thick brane DCE in $f(R,T)$ gravity}

The first step in calculating the DCE is to obtain the modal fraction of the system (\ref{EnergyD}). To find the modal fraction, we substitute the warp function (\ref{Afunction}), the matter field solution (\ref{MatterF}), and the interaction (\ref{Pot}) in terms of the extra dimension in Fourier's transform (\ref{FourierT1}). Posteriorly, considering the solution of $G(\omega)$, one obtains, after an extensive calculation, the modal fraction of the brane. In this case, the modal fraction is
\begin{align}\label{modalf2}
    g(\alpha,\,\sigma;\, \omega)=&\frac{105\pi[2\alpha\sigma^2\omega+(3+5\alpha)\omega^3]^2}{16\sigma^5[35\alpha^2+28\alpha(3+5\alpha)\sigma+20(3+5\alpha)^2\sigma^2]}\bigg[-1+\cosh\bigg(\frac{\pi\omega}{\sigma}\bigg)\bigg]^{-1}
\end{align}
Therefore, one can note that the DCE will be changed when the metric fluctuation and brane thickness varies.

\begin{figure}[ht!]
    \centering
    \includegraphics[height=7cm,width=8cm]{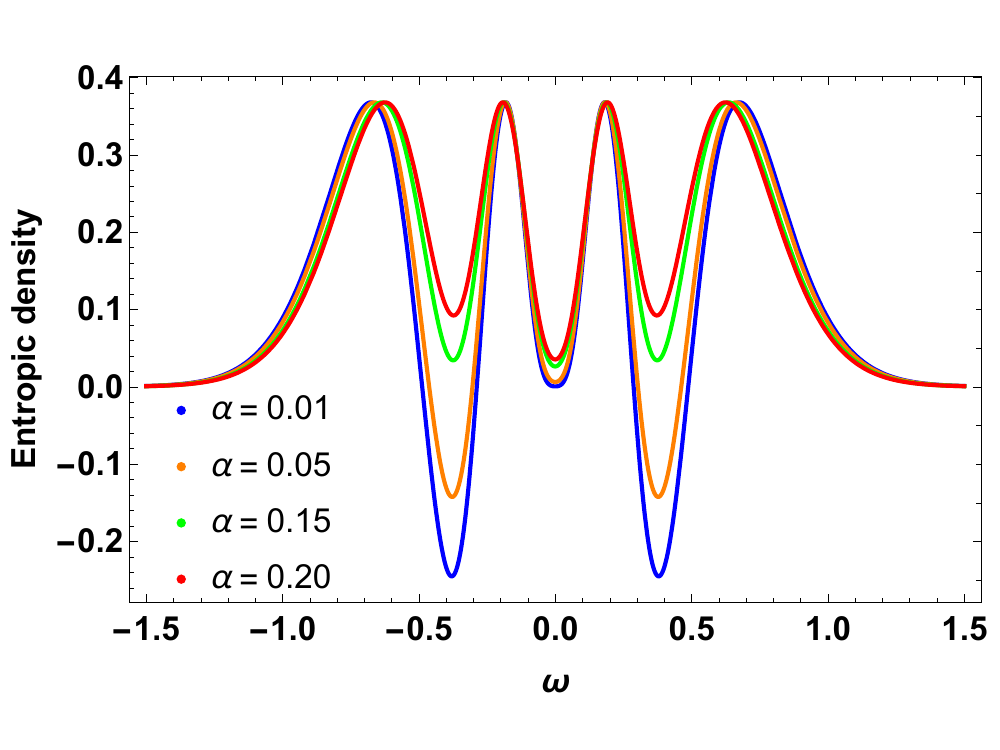}
    \includegraphics[height=7cm,width=8cm]{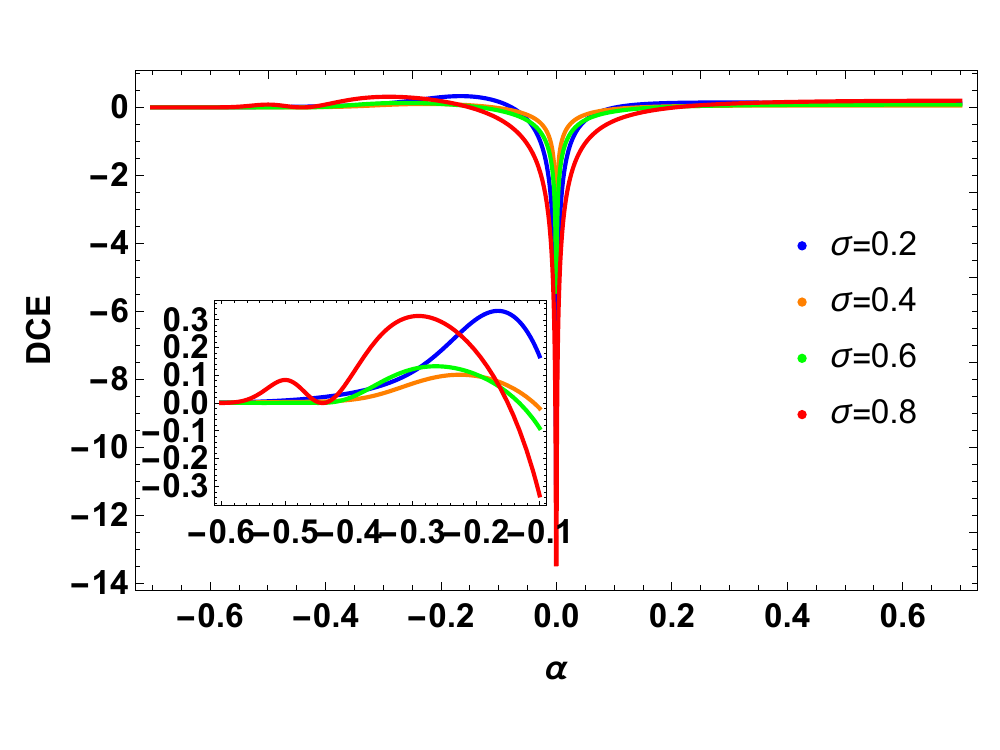}    
    \vspace{-0.3cm}
    \begin{center}
      \hspace{0.3cm}  (a) \hspace{8cm} (b)
    \end{center}
    \vspace{-0.3cm}
    \caption{(a) Entropic density of the DCE for several values of metric fluctuation. (b) DCE of the braneworld in terms of the fluctuation.}
    \label{fig5}
\end{figure}

Using the modal fraction (\ref{modalf2}), we calculate the DCE (\ref{DCE}) from the brane. To perform this calculation, a numerical investigation of the solution of the integral (\ref{DCE}) is considered. The numerical result of  DCE is shown in Fig. \ref{fig5}(b). Meanwhile, the entropic density associated with numerical solutions is found in Fig. \ref{fig5}(a).

Interesting results arise when analyzing the DCE of the brane in $f(R,T)$ gravity induced by metric fluctuations, i.e., DCE reaches maximum values when $\sigma$ increases [see Fig. \ref{fig4}(b)]. Furthermore, there is a critical point of the DCE at $\alpha=0$ regardless of the thickness of the brane. That suggests that the most likely structures appear when the metric fluctuations are zero, i.e., in the usual theory without gravity modifications. However, other local critical points occur in $\alpha=0.04$. Thereby, we note, numerically, that the most likely and stable field configurations are kink-like and emerge when $\vert\alpha\vert= 0.4$ and $\sigma=0.8$, i. e., in the modified gravity scenario.

\section{Final remarks}

In this work, one studies the influence of metric fluctuations in a braneworld scenario. We noted that when applying the metric quantum perturbations in the Einstein-Hilbert action coupled to the matter field, a modified gravity theory is obtained and described by the function $f(R, T)=-\frac{1}{4} (1-\alpha)R+\frac{1}{2}\alpha T$. This result is interesting once it allows us to recover the usual case, i.e., the Einstein-Hilbert theory, when the metric fluctuations are null, i.e., $\alpha\to 0$.

Considering the results obtained by the perturbative approach, we build a braneworld in $f(R,T)$ gravity. In this scenario, one notes that the vacuum states depend on the quantum fluctuations of the metric. Consequently, due to these dependencies, the asymptotic value of the matter field is changed. Furthermore, it is possible to notice that when the fluctuation reaches maximum values, the domain wall will no longer exist. That is because when $\alpha\to 1$, the vacuum expected value is null. Meantime, for small metric fluctuations, i.e., $\vert\alpha\vert<1$, the domain walls arise. So that when $\vert\alpha\vert\ll 1$, the fluctuation effects will feel far from the brane. These results influence the brane energy, where critical points increase their intensity around $y=0$ when $\vert\alpha\vert\ll 1$.

Finally, we use the differential configurational entropy formalism to study the most likely, and stable matter field configurations. In this analysis, some attractive results emerge, i.e., the DCE reaches absolute maximums (or minimum) at $\alpha=0$ independent of the brane thickness. This critical point is indicated in $\alpha=0$ [Figs. \ref{fig4}(a) and (b)], regardless of brane thickness, suggests that the most likely structures appear when the metric fluctuations are zero, i.e., when we recover the usual theory. However, another local critical point occurs at $\alpha\simeq 0.04$, indicating configurations more likely and stable are kink-like structures and appear in the modified gravity scenario when $\vert\alpha\vert\simeq 0.4$ and $\sigma=0.8$.

Although we have developed a theoretical discussion, one can confront these results experimentally. That is possible because the modified gravity theories presented can be verified through cosmography \cite{SCapozziello1,SCapozziello2}. This analysis is feasible, given that the cosmological principle postulates a single factor as the primary degree of freedom governing the universe \cite{SCapozziello1,SCapozziello2}, i.e., 
\begin{align}\label{ce1}
    a(t)=1+\sum_{k=1}^{\infty}\frac{1}{k!}\frac{d^{k}a}{dt^k}\bigg\vert_{t=t_0}(t-t_0)^k.
\end{align}
Using $a(t)$, one can estimate the Hubble, deceleration, jerk, and snap parameters, respectively, as
\begin{align}\label{ce2}
    H(t)=\frac{1}{a}\frac{da}{dt}, \hspace{0.5cm} q(t)=-\frac{1}{aH^2}\frac{d^2a}{dt^2}, \hspace{0.5cm} j(t)=\frac{1}{aH^3}\frac{d^3a}{dt^3}, \hspace{0.5cm} \text{and} \hspace{0.5cm} s(t)=\frac{1}{aH^4}\frac{d^4a}{dt^4}.
\end{align}
In this form, it is possible to confront the data obtained in our study with observational data. That is feasible because it allows for the correlation of cosmological parameters with the parameter of modified gravity arising from quantum fluctuation.

A future perspective of this study is to understand how these fluctuations influence cosmological objects and their properties. We hope to perform this study soon.

\section*{ACKNOWLEDGMENT}
The authors thank the Brazilian research agencies, i.e., CNPq, CAPES, and FAPEMA, for financial support. C. A. S. Almeida thanks the Conselho Nacional de Desenvolvimento Cientifico e Tecnologico (CNPq), grant no 309553/2021-0. Furthermore, C. A. S. Almeida is grateful to the Print-UFC CAPES program, project number  88887.837980/2023-00, and acknowledges the Department of Physics and Astronomy at Tufts University for hospitality. F. C. E. Lima is supported by FAPEMA BPD-05892/23 and grateful to the Department of Physics from the Universidade Federal do Ceará (UFC) for hospitality and support.

\section*{CONFLICTS OF INTEREST/COMPETING INTEREST}

All the authors declared that there is no conflict of interest in this manuscript.

\section*{DATA AVAILABILITY}

No data was used for the research described in this article

\end{document}